\documentstyle[emulateapj,onecolfloatx]{article}

\newcommand{\beq}{\begin{equation}}
\newcommand{\eeq}{\end{equation}}

\def\beqa{\begin{eqnarray}}
\def\eeqa{\end{eqnarray}}

\def\lb{\langle}
\def\rb{\rangle}

\lefthead{Peebles}
\righthead{The Void Phenomenon}

\begin{document}

\twocolumn[

\submitted{}

\title{The Void Phenomenon}
\author{P. J. E. Peebles}
\affil{Joseph Henry Laboratories, Princeton University,
Princeton, NJ 08544}

\begin{abstract}
Advances in theoretical ideas on how galaxies formed have not
been strongly influenced by the advances in observations
of what might be in the voids between the concentrations of
ordinary optically selected galaxies. The theory and observations
are maturing, and the search for a reconciliation offers a
promising opportunity to improve our understanding of cosmic
evolution. I comment on the development of this situation and
present an update of a nearest neighbor measure of the void 
phenomenon that may be of use in evaluating theories of galaxy
formation. 
\end{abstract} 
\vspace{3 mm}\
]

\section{Introduction}
Voids between the regions occupied by normal textbook galaxies
contain few observable galaxies of stars or gas clouds, and,
apart from a tendency to greater gas content and star formation
rates, objects observed in and near voids seem to be close to a fair
sample of the cosmic mix. This striking effect will be termed the
void phenomenon. A numerical definition, based on nearest
neighbor statistics, is summarized in \S 3.4. 

The void phenomenon may be related to the preferences of
early-type galaxies for high density regions and
of galaxies rich in gas and star formation for the
edges of voids. This morphology-density correlation follows in a
natural way from the biased galaxy formation picture in the
Gaussian adiabatic cold dark matter (CDM) model. Perhaps the
biased formation extends to the voids, where the morphological
mix swings to favor  
dark galaxies. But if this were so the shift to the void mix
would have to be close to discontinuous (\S 2.2), in remarkable 
contrast to the observed relatively slow variation of
morphological mix with ambient density in regions occupied by
visible galaxies. From continuity one might have thought the
more likely picture is that gravity has emptied the 
voids of mass as well as galaxies. This does not happen in the
CDM model, however. Simulations show, between the
concentrations of large dark mass halos, clumps of mass that
seem to be capable of developing into void objects observable as 
clumps of stars or gas, contrary to what is observed. 

The apparent discrepancy has not attracted much attention since
the introduction of the biased galaxy formation picture fifteen
years ago. This is partly because we do not have an established
theory of how mass concentrations become observable (though we do
have some guidance from the observations of galaxies at ambient
conditions that do not seem to be very different from the voids;
\S 4.1). A contributing factor is  
that the theoretical community has not settled on a standard
numerical measure of the void phenomenon. A nearest neighbor
statistic is commonly used in the observational community. This 
is discussed in \S 3, and \S 3.4 summarizes an update
of the nearest neighbor statistic that might be applied to
realizations of galaxy formation models.

The main points of this paper are summarized in \S 5.1. Section
5.2 offers comments on the history of the void phenomenon as an
example of the complex and occasionally weak interaction of
theory and practice in a developing subject like cosmology. The
discussion of the observational situation in \S 2 and \S 3 and of 
the theoretical situation in \S 4 illustrates the unusually
lengthy duration of modest interaction between theoretical and
observational ideas about voids, and the resulting opportunity
now to learn something new.  

\section{Voids and Void Objects}

\subsection{Voids}

\cite{Rood81} (1981) gives an excellent picture of the state of
observational studies of galaxy clustering two decades
ago. He remarks that 20 years earlier \cite{Mayall60} (1960,
fig. 3) had found that in the direction of the Coma cluster far
more galaxies are in the cluster than the foreground. Redshift
surveys to larger angular distances from the cluster
(\cite{TG76} 1976; \cite{CR76} 1976; \cite{GT78} 1978)
show \cite{Mayall60} had observed foreground 
voids\footnote{The statement in \cite{Pee93} (1993) that
\cite{KOSS81} (1981) named these regions voids is wrong; 
Kirshner et al. clearly state the prior discussions.} with 
radii\footnote{Hubble's constant is $H_o=100h$ km s$^{-1}$
Mpc$^{-1}$.}  $\ga 20h^{-1}$~Mpc. \cite{Ei78} (1978)
independently identified voids --- holes in their terminology --- 
in the galaxy distribution in the southern galactic hemisphere.

Other surveys of galaxy types --- dwarf, low luminosity,
irregular, low surface brightness, star-forming, or IRAS --- that
are found to respect common voids include \cite{Markarian83}
(1983), \cite{Thuan87} (1987), \cite{Eder89} (1989), \cite{Binggeli90}
(1990), \cite{LSB93} (1993), \cite{BCG96} (1996), \cite{Kuhn97}  
(1997), \cite{Schombert97} (1997), \cite{Popescu97} (1997),
\cite{ELG00} (2000), and  \cite{Piran00} (2000). 

The preference of star-forming galaxies for the edges of voids
certainly agrees with the biased galaxy formation picture, as
Salzer has consistently noted (\cite{Salzer90} 1990; \cite{ELG00}
2000; and references therein). Salzer also emphasizes the
similar large-scale structures. Figure~3 in \cite{ELG00} (2000)
clearly illustrates voids defined by both ordinary and
emission-line galaxies.  

Gas clouds also avoid voids. This applies to gas detected 
in HI 21-cm emission (\cite{HIb91} 1991; \cite{HIa92} 1992;
\cite{HIc96} 1996; \cite{Zwaan97} 1997) and gas detected at
somewhat lower surface densities as Lyman-limit or MgII quasar
absorption line systems: when the redshift allows the test a
galaxy generally is observed close to the line of sight and near
the redshift of each absorption line system (\cite{Bergeron91}
1991; \cite{Steidel94} 1994; \cite{Lanzetta95} 1995).   
\cite{Shull96} (1996) show that gas clouds detected as very low
surface density Lyman~$\alpha$ absorbers, with HI surface density 
$\sim 10^{13}$~cm$^{-2}$, avoid dense galaxy 
concentrations. A visual impression is that they also avoid the
voids, but that awaits a numerical test, perhaps along the lines
of the nearest neighbor distribution (\S 3). 

\subsection{Void Boundaries}

The strong clustering of galaxies on small scales must open
low density regions. In a clustering hierarchy model
that fits the two- through four-point correlation functions of
optically selected galaxies (\cite{SP78} 1978) the void sizes are
comparable to what is observed (\cite{Soneira78} 1978;
\cite{Vett85} 1985). The discovery from redshift maps 
is that the void boundaries tend to be rather smooth, 
and defined by galaxies with a broad range of
luminosities (\cite{bridges81} 1981; \cite{deL86} 1986). 

\subsection{Void Objects}

Another aspect of the void phenomenon is that isolated galaxies 
do not seem to be particularly unusual, apart
from the gas content. Thus \cite{HIc96} (1996) conclude that ``the
void galaxies [in Bo\"otes] seem to be unaware of the fact that
they exist in a huge underdense region.'' In an important
advance \cite{GroginGeller00} (2000) find evidence that the
relative velocities of void galaxy pairs at
projected separation $<115h^{-1}$~kpc tend to be lower than the
relative velocities at near cosmic ambient density. 
Pairs of galaxies at separation $\sim 100h^{-1}$~kpc typically
have other neighbors whose massive halos add to the relative
velocity, however, so the implication for the correlation of halo
mass with ambient density will require further discussion. 

\subsection{Voids and the Morphology-Density Correlation}

The mix of morphological types correlates with
ambient density (\cite{Hubble36} 1936; \cite{Dressler80} 1980;
\cite{Postman84} 1984). The new aspect of the void phenomenon is
that if there is a special class of void galaxies the shift to
the void mix when the ambient density falls below some fraction
of the cosmic mean has to be close to discontinuous.

This point seems to have been first made by \cite{KOSS81} (1981), 
who note that if the voids were missing galaxies but not mass
one might expect low luminosity galaxies are more common relative
to $L\sim L_\ast$ galaxies in voids. But they conclude that if
this were so the void galaxies would 
have to be several magnitudes fainter than typical optically
selected galaxies, because these hypothetical faint galaxies are
not observed. This agrees with the CfA
(Center for Astrophysics) galaxy maps in Figures~2a and~2d of
\cite{CfA82} (1982): the distributions of the high and low
luminosity galaxies are strikingly similar. The same effect is
seen in the extension of the CfA survey (\cite{deL86} 1986).

The morphology-density correlation includes the preference of
star-forming galaxies for the edges of voids 
(\cite{Markarian83} 1983; \cite{Salzer89} 1989). But Figure~3 
in \cite{GroginGeller00} (2000) shows how
subtle is the variation of the mix of observed types with 
ambient density when the density is comparable to
the cosmic mean. It can be compared to the very different 
mix within voids, if there is a dark void population. 

\begin{figure}
\plotone{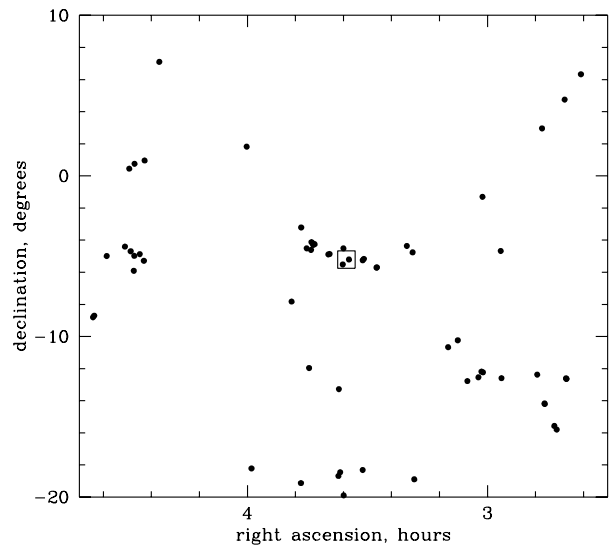}
\caption{Map of Optical Redshift Survey galaxies (filled circles)
in a slice in redshift space 600~km~s$^{-1}$ deep centered on
SBS~0335-052 (open box). The field is $23.5h^{-1}$~Mpc wide by 
$21.3h^{-1}$~Mpc high.} 
\end{figure}

\subsection{The Most Extraordinary Objects}

Extremely unusual galaxies are worth special consideration. The
statistics of their environments are insecure, by definition, but
the discovery of a few cases well inside voids would be
influential. Two examples of quite unusual objects --- that happen
not to be in voids --- may illustrate the situation.   

A systematic program of study of the remarkable blue compact
galaxy SBS~0335-052 (\cite{Thuan97} 1997;
\cite{Vanzi00} 2000; \cite{Pustilnik00}
2000; and references therein) shows it has an
extended HI cloud with quite low heavy element abundances along
the lines of sight to star-forming regions, quite young star
populations, and a large mass-to-light ratio. It has many of the
properties one might look for in a young galaxy at low redshift,
except its position. The projected 
distance to the large spiral galaxy NGC~1376 is 150~kpc, at close 
to the same redshift (\cite{Pustilnik00} 2000). Figure~1 shows
the position of SBS~0335-052 relative to 
the galaxies in the Optical Redshift Survey
(\cite{Santiago95} 1995)\footnote{ 
http://www.astro.princeton.edu/~strauss/ors/index.html}  
with heliocentric redshifts in the range $4043\pm 300$~km~s$^{-1}$
centered on SBS~0335-052. This unusual object is not in a dense
region. More important of the present purpose, it is not in a
void. The former agrees with the familiar behavior of
star-forming galaxies. The latter does not naturally agree with
the biased galaxy formation picture.  

A second example is DDO~154. The large ratio of dark to luminous
mass led \cite{Carignan88} (1988) to term it a ``dark''
galaxy. It is not far from the Local Group, at supergalactic
coordinates SGL$=90^\circ$, SGB$=+7^\circ$, in a continuation of
the local sheet of galaxies rather than the nearby voids.

A void population would be expected to have a range of
properties, some with enough stars or gas to be detectable. 
SBS~0335-052 and DDO~154 are strange enough to motivate the
thought that they are unusual members of this hypothetical void 
population, not extremes from known populations.
But that does not agree with their positions, near galaxies. 

These are just two examples. Systematic studies of environments
of unusual galaxies, particularly early types, will be followed
with interest. 

\section{Statistical Measures}

The tendency of less ordinary types of galaxies to avoid the
voids defined by ordinary spirals has been probed by void
probability functions (\cite{White79} 1979; \cite{Vogeley94}
1994; \cite{Piran00} 2000; and references therein), two-point
correlation functions, and nearest neighbor distances. To keep
this discussion somewhat limited I discuss only the latter
two, that have complementary features.  

A particularly useful feature of correlation functions is 
the simple relation between the measurable angular
function and the wanted spatial function. There is no
simple relation between nearest neighbor distances in real space 
and the measurable nearest distances in projected angular
distributions or in redshift space. But as discussed next, the
interpretation of correlation functions as a probe of the void
phenomenon is not straightforward. The rest of this paper
accordingly uses nearest neighbor statistics. 

\subsection{Two-Point Statistics}

One sees in \cite{Markarian83}'s (1983) maps the tendency of
Markarian galaxies to avoid both dense regions and voids in
the distribution of ordinary galaxies. The former means that
on small scales the Markarian correlation function, $\xi _{MM}$,
and the Markarian-optically selected galaxy cross correlation
function, $\xi _{Mg}$, are significantly less than the
galaxy-galaxy function, $\xi _{gg}$. If at larger separations  
$\xi _{MM}$, $\xi _{Mg}$, and $\xi _{gg}$ had similar values,
because both galaxy types trace the same large-scale structure, 
the slopes of $\xi _{MM}$ and $\xi _{Mg}$ would be shallower
than $\xi _{gg}$. This would not mean Markarian galaxies tend to
occupy the voids defined by ordinary galaxies, of course.
To the contrary, the visual impression from \cite{Markarian83}'s
maps and the evidence from \cite{Markarian83}'s nearest neighbor
statistic is that Markarian galaxies respect the voids. 

The correlation functions for early- and late-type galaxies in 
\cite{twopoint96} (1996 Figs.~10 and~11) behave as just described:
at small separations later types have distinctly smaller
correlation functions, while at separation 
$hr\sim 10$~Mpc the functions have similar values. This 
certainly demonstrates the small-scale morphology-density
correlation. One sees from the example of Markarian galaxies that
it need not imply later-type galaxies tend to occupy the
voids defined by earlier types.  

If at $hr\ga 10$~Mpc the correlation function for an unusual 
type of galaxy were unusually large it could be a signature of
void galaxies. But because one can trace 
connections among occupied regions over quite large distances one
would also have to consider the way the distributions of objects
within occupied regions contribute to the abundances of galaxy
pairs at large separations. 

\begin{figure}
\plotone{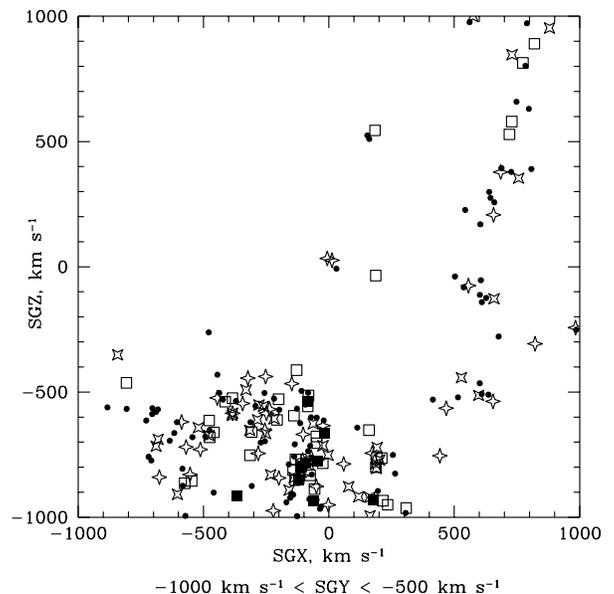}
\caption{Map of Optical Redshift Survey galaxies in a slice
in redshift space. The normal to the slice points in the direction
of the Virgo Cluster, from the opposite side of the Milky Way. 
Filled squares are elliptical galaxies, open
squares S0s, crosses Sa to Sb-c, plus signs later
spiral types, and filled circles dwarfs and irregulars.}
\end{figure}

\subsection{Nearest Neighbor Statistics}

This measure also is affected by the morphology-density
correlation, but it may reveal void galaxies through a tail in
the distribution of nearest neighbor distances. 

The test uses two types of objects.
The $o$-type are reference ordinary galaxies, or the proposed 
equivalent in a simulation. The $t$-type are
unusual test objects that may have tended to form in the unusual
conditions in voids between concentrations of $o$-types.
In analyses of observations the $t$-types may be galaxies ---
dwarf, irregular, compact, or low surface brightness --- or gas
clouds. In analyses of simulations the $t$-types would be mass
concentrations that are not expected to develop into ordinary
galaxies but seem  to be capable of forming observable
concentrations of stars or gas.

The distance from a $t$-type to the nearest $o$-type object is
$D_{to}$, and the distance from an $o$-type to the nearest
neighboring $o$-type is $D_{oo}$. The probability distribution of
$D_{to}$ 
depends on the number density of the reference $o$-type objects,
but we have a control from the distribution of $D_{oo}$. If the
two types were randomly selected from the same underlying
population, and the two selection probabilities
differed by a constant factor, 
the distributions of $D_{to}$ and $D_{oo}$ would be the same
within the noise, even if the selection functions varied with 
position. If the $t$-types tended to be outside the
concentrations of $o$-types it would produce a tail of large
values in the distribution of $D_{to}$ relative to $D_{oo}$.

All applications of this probe have used nearest neighbor
distances in redshift space, which can be much larger than in
real space (\S 3.3). The effect can be analyzed by using as
separate variables the transverse and redshift separations,
but this will be left for future work.  

Early examples of this probe are in \cite{SP77} (1977), who used
it to test and argue against the idea that there is a spatially
homogeneous population of field galaxies, and
\cite{Markarian83} (1983), who introduced it to the analysis of
voids.\footnote{This study was 
used in \S 3.1 to illustrate the problematic interpretation of 
correlation functions as a measure of voids. But
\cite{Markarian83} (1983) used nearest neighbor statistics,
not correlation functions.}   

Nearest neighbor statistics have been applied to a considerable
variety of candidate void objects 
(\cite{Eder89} 1989; \cite{Salzer90} 1990; \cite{LSB93} 1993;
\cite{Pustilnik95} 1995; \cite{BCG96} 1996; \cite{ELG00} 2000).
In some cases the 
nearest reference ordinary galaxy is about equally close
to test and ordinary galaxies. In others the test sample have the
more distant nearest neighbors. The difference is not very
large, however. Thus one finds general agreement that
all these classes of objects avoid the central parts of voids,
but mixed opinions on consistency with the picture of biased
galaxy formation. Two new applications of the statistic
may help clarify the situation.

\begin{figure}[t]
\plotone{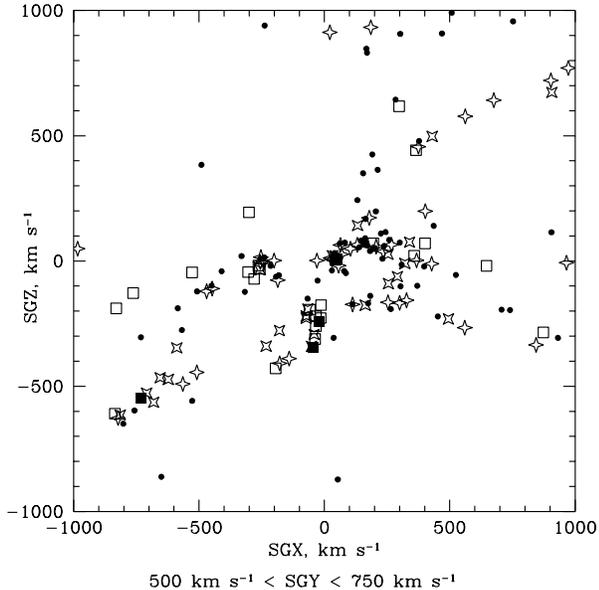}
\caption{Map of ORS galaxies in a slice
in redshift space on the same side of the Milky Way as the 
Virgo Cluster.}
\end{figure}

\begin{figure}[t]
\plotone{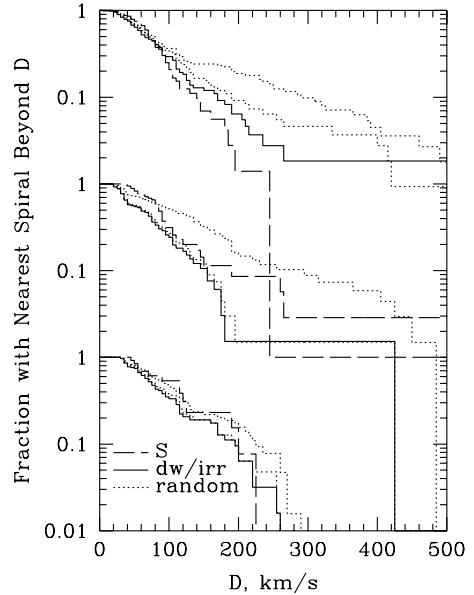}
\caption{Distributions of distances $D$ in redshift space of the
nearest spiral neighbors of spirals (broken lines) 
and of dwarfs and irregulars (solid lines). The upper dotted
lines show the effect of randomly shifting one in three of the
dwarfs and irregulars, the lower dotted lines the effect of
randomly shifting one in ten. The redshift ranges are  
$200 < cz < 500$ km~s$^{-1}$ in the bottom plot, 
$500 < cz < 750$ km~s$^{-1}$ in the middle, and 
$750 < cz < 1000$ km~s$^{-1}$ in the top.} 
\end{figure}

\subsection{New Examples of the Nearest Neighbor Statistic}

Redshift samples have improved, so an update of the nearest
neighbor statistic is worthwhile. The example in \S 3.3.1 
uses dwarfs plus irregulars as test galaxies, and in \S 3.3.2 low
surface brightness galaxies.  

Both examples use the spirals in the Optical
Redshift Survey (ORS; \cite{Santiago95} 1995) as reference
ordinary galaxies. At low redshift the ORS broadly samples 
the luminosity function, and it has useful information on
morphology. The authors caution that completeness as a function
of apparent magnitude and surface brightness varies across 
the sky; the selection may be particularly inhomogeneous
for dwarfs and irregulars. The latter need not affect the 
relative distributions of $D_{to}$ and $D_{oo}$ if the ordinary
galaxies are homogeneously sampled. And inhomogeneous sampling
would not seem likely to mask the signature of void galaxies. 

\begin{table*}
\begin{center}
\caption{\hspace{2.75in}Nearest Neighbor Parameters}
\begin{tabular}{lrrrrrrrrrr}
\vspace{-5pt}\\
\tableline
\tableline
\vspace{-8pt}\\
Test Sample\quad & $cz_{\rm min}$\tablenotemark{a}
	& $cz_{\rm max}$\tablenotemark{a} & $N_o$ & $N_t$ & 
	$n_o$\tablenotemark{b} & $n_t$\tablenotemark{b} 
	& $H_oD_{h}$\tablenotemark{a} & 
	$H_o\langle D_{oo}\rangle$\tablenotemark{a} & 
	$\langle D_{to}\rangle /\langle D_{oo}\rangle$ &
	$\langle D_{ro}\rangle /\langle D_{oo}\rangle$\\
\vspace{-9pt}\\
\tableline
\vspace{-9pt}\\
Dw/Irr & 200\quad & 500\quad  & 20 & 63 & 0.08 & 0.26 & 11 & $110\pm 17$&
	$0.83\pm 0.14$ & $1.43 \pm 0.23$ \\
Dw/Irr & 500\quad  & 750\quad  & 51 & 66 & 0.08 & 0.11 & 11 & $107\pm 16$&
	$0.69\pm 0.12$ & $1.99 \pm 0.32$ \\
Dw/Irr & 750\quad  & 1000\quad  & 100 & 109 & 0.08 & 0.09 & 11 & $78\pm 5$&
	$1.10\pm 0.12$ & $3.09 \pm 0.26$ \\
LSBa & 200\quad & 3000\quad & 153 & 27 & 0.02 & 0.003 & 38 & $133\pm 10$&
	$0.92\pm 0.14$ & $ 2.51 \pm 0.30$ \\
LSBb & 200\quad & 3000\quad & 70 & 43 & 0.02 & 0.01 & 32 & $115\pm 10$ & 
	$0.88\pm 0.12$ & $3.19\pm 0.36$ \\
LSBa & 3000\quad & 6000\quad & 160 & 43 & 0.003 & 0.0008 
	& 180 & $284\pm 17$& $1.10\pm 0.12$ & $1.68 \pm 0.15$ \\
LSBb & 3000\quad & 6000\quad & 80 & 60 & 0.004 & 0.003 & 140 
	& $270\pm 20$ & $1.03\pm 0.10$ & $1.81 \pm 0.17$ \\
LSBa & 6000\quad  & 9000\quad & 106 & 42 & 0.0007 & 0.0003 
	& 560 & $490\pm 40$ & $1.10\pm 0.12$ & $ 1.39 \pm 0.17$ \\
\tableline
\multicolumn{2}{l}{\footnotesize $^a$Unit km~s$^{-1}$} & 
\multicolumn{2}{l}{\footnotesize $^b$Unit $h^3$Mpc$^{-3}$} \\
\end{tabular}
\end{center}
\end{table*}

\subsubsection{Redshift Maps}

Figures 2 and 3 show maps of the ORS galaxies in nearby narrow
slices in redshift space (corrected to the Local Group by
adding to the heliocentric redshift 300 km~s$^{-1}$ toward $l =
90^\circ$ $b=0$). The normals point to supergalactic coordinates
$SGB=0$,  
$SGL = 90^\circ$, roughly toward the Virgo Cluster. The slice in
Figure~2 is on the opposite side of the Milky Way from the Virgo
Cluster, the slice in Figure~3 on the same side. 
Galaxies closer than $6^\circ$ from the Virgo Cluster are removed.
The slice toward Virgo is thinner, because this side is more
crowded, and it is seriously distorted by
virgocentric flow, but early and late types may be similarly
affected by large-scale streaming. Most parts of the 
slices are at high galactic latitude, but the upper right corner
in Figure~2 and the lower left corner of Figure~3 dip to
$|b|<30^\circ$, where the ORS may be significantly less complete.  

The filled squares are elliptical galaxies (Burstein numerical
morphological type BNMT$\le 12$, de Vaucouleurs type $T\le -5$; 
\cite{Willik97} 1997, Table~7).
The open squares are S0s (cut at BNMT$\le 112$, $T\le 0$), 
the crosses Sa through Sb-c 
(cut at BNMT$\le 152$, $T\le 4$), 
the plus signs later type spirals 
(cut at BNMT$\le 182$, $T\le 7$), 
and the filled circles all the dwarfs, irregulars, and other
categories.

It is difficult to make out much difference in clustering
properties of the different morphological types in Figures~2
and~3. It is easy to find nearly empty regions. The former is
at least in part due to the absence of rich clusters. The latter
may also be affected by the very limited samples, it agrees
with the many other examples reviewed in \S 2 and \S 3.2. 

\subsubsection{Dwarfs and Irregulars}

Results of the nearest neighbor test applied to the ORS Dw/Irr
galaxies (filled circles in Figs.~2 and~3) and spiral
galaxies (crosses and plus signs) are shown in Figure~4 and Table~1.  

Pairs of ORS galaxies closer than 2~arc min and at low
redshift tend to have the same or quite similar 
redshifts, a sign they may be the same galaxy, so one
is discarded. I cut the galaxies at galactic
latitude $|b|<30^\circ$ and within $6^\circ$ of the
Virgo cluster ($\alpha = 186.6^\circ$, $\delta = 13.2^\circ$).

The subdivision in the three bins in redshift reduces the effect
of variable completeness as a function of redshift, and it offers
a test of reproducibility. 

For each galaxy, $o$ (spiral) or $t$ (Dw/Irr), in each redshift
bin, Figure~4 shows the distribution of distances in redshift
space (relative to the Local Group) to the nearest
$o$-type at any redshift greater than 200~km~s$^{-1}$ and within 
the cuts in angular position. The plots are normalized 
cumulative distributions. 

In the two lowest redshift bins the Dw/Irr
galaxies tend to have the closer spiral neighbor; in
the highest redshift bin the spirals tend to have the closer
spiral neighbor. The difference may be related to the variation
in the range of absolute magnitudes sampled in each redshift bin
(reflected in the variation in the ratio of numbers $N_t$ and
$N_o$ of Dw/Irr and spirals). Or it may only be noise, for the
differences in the distributions are small.  

Table~1 shows the means $\lb D_{oo}\rb$ of the distances of the
nearest spiral neighbors of spirals, and the ratios
$\langle D_{to}\rangle /\langle D_{oo}\rangle$ of the  
means of nearest neighbor distances of Dw/Irr
and of spiral galaxies. The standard deviations
assume each distance is statistically independent. (If
two $o$-types are each other's nearest neighbor the distance
counts once.) The differences of the distributions in Figure~4 
may appear more significant than the differences of the 
ratios $\langle D_{to}\rangle /\langle D_{oo}\rangle$ from unity, 
because the cumulative distributions over-emphasize the noise.

We can estimate the nearest neighbor distance among spirals in 
real space from the observation that the small-scale 
distribution of optically selected galaxies approximates a 
scale-invariant clustering hierarchy, or fractal, with two-point
correlation function $\xi =(r_o/r)^\gamma$. The distance $r_h$ at
which a spiral has on average one spiral neighbor satisfies
\beq
	n_o\int _0^{r_h} \xi\, d^3r = 
	{4\pi n_or_h^{3 - \gamma}r_o^\gamma \over 3 - \gamma} = 1,
\eeq
if the mean density $n_o$ is large enough that $r_h\ll r_o$.
Since the small-scale distribution is close to scale-invariant
the mean $D_h$ of the nearest neighbor distance differs from $r_h$
by a fixed factor. Since the factor is not likely to be greatly
different from unity a useful estimate of the mean real distance
from a spiral to the nearest spiral is 
\beq
D_h\sim\left( 3-\gamma\over 4\pi n_o r_o^\gamma\right)
	^{1/(3 - \gamma)}.
\label{eq:Dh}
\eeq
The table lists $H_oD_h$ for $hr_o=5$~Mpc and $\gamma =1.77$.

The small-scale relative peculiar velocity in the field 
is an order of magnitude larger than $H_oD_h$ in the ORS at 
$cz<3000$ km~s$^{-1}$. This means the ratio 
$\lb D_{to}\rb /\lb D_{oo}\rb$ is determined by peculiar
velocities; it does not tell us whether Dw/Irr or
spiral galaxies tend to have the closer spiral neighbors in 
position space. 

The $D_{to}$ could be reduced by accidental cancellation of
differences in the distributions of relative velocities and
positions, but that seems unlikely. Thus I conclude   

(1) the distributions of peculiar velocities of Dw/Irr and
spiral galaxies relative to nearby spirals are quite similar, and

(2) the mean distance in position space from a Dw/Irr to the
nearest spiral is less than about $1h^{-1}$~Mpc.  

The second result depends on the number density of spirals, here
$n_o\sim 0.1h^3$ Mpc$^{-3}$. But, since most spirals avoid the
voids, if $n_o$ were decreased by cutting the sample at larger
luminosity or circular velocity, the Dw/Irr's would still tend 
to be within $1h^{-1}$~kpc of the concentrations of the sample
spiral galaxies.

The last column in the table shows the result of moving each of
the Dw/Irr galaxies to a randomly chosen position, with constant
probability per unit solid angle within the cuts in galactic
latitude and distance from the Virgo Cluster, and at
uniformly probability distribution in $z^3$ within the 
redshift bin for the galaxy. The increase in nearest neighbor
distances is significant, though not all that large. 

The separate test illustrated by the upper dotted lines in
Figure~4 shows the result of shifting to a random position one in
three of the Dw/Irr (selecting every third object in the catalog
list for a new position), leaving the rest at their catalog
positions. The lower dotted lines show the effect of randomly
repositioning one in ten. At one in three the resulting tail in
the nearest neighbor distances is marginal in the shallowest
sample, pronounced in the two deeper ones. It follows that
at the catalog positions fewer than one in three of the Dw/Irr's
could be members of a homogeneously distributed population, for
otherwise the distributions of $D_{to}$ for the other galaxies
could not be tight enough to match the observed distribution of
all  $D_{to}$. When one in ten is randomly shifted the tail is
pronounced only in the deepest sample. Thus I conclude

(3) fewer than about one in ten of the Dw/Irr is in a
statistically homogeneous randomly distributed population.

\begin{figure}[t]
\plotone{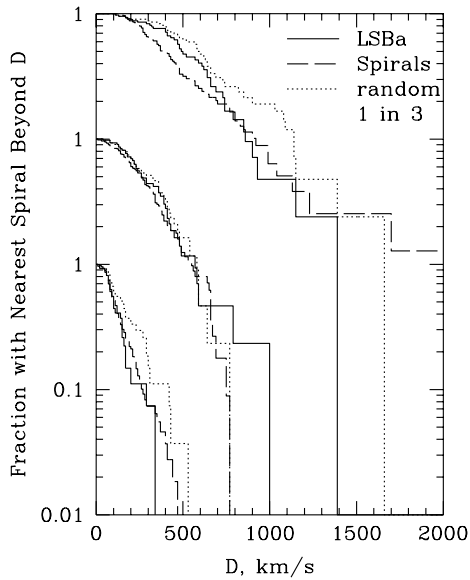}
\caption{Distributions of distances of the nearest spiral
neighbors of spirals and of \cite{LSBSample92} (1992) 
LSB galaxies. The dotted lines show the effect of randomly
shifting one in three of the LSB galaxies. The redshift
ranges are $200 < cz < 3000$ km~s$^{-1}$ 
in the bottom plot, $3000 < cz < 6000$ km~s$^{-1}$
in the middle, and $6000 < cz < 9000$ km~s$^{-1}$
in the top.}
\end{figure}

\begin{figure}
\plotone{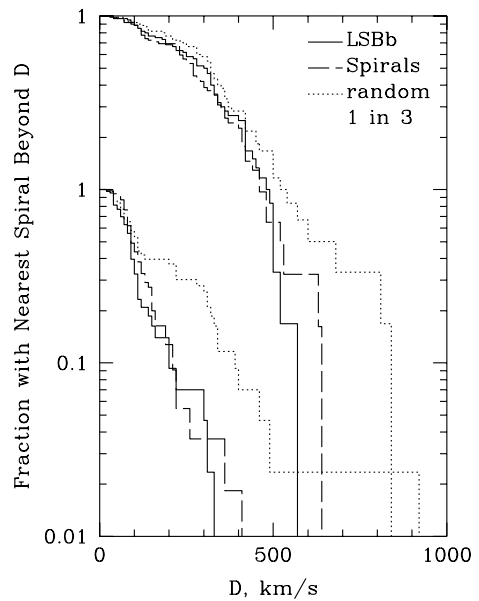}
\caption{The same as Fig. 5 for the \cite{Impey96}
(1996) LSB galaxies at redshift
ranges $200 < cz < 3000$ km~s$^{-1}$ 
in the bottom plot and $3000 < cz < 6000$ km~s$^{-1}$
in the top.}
\end{figure}

\subsubsection{Low Surface Brightness Galaxies}

The \cite{LSBSample92} (1992; here LSBa) and 
\cite{Impey96} (1996; here LSBb)
low surface brightness galaxies are analyzed
separately to check reproducibility. The reference ordinary
galaxies are the ORS spirals used in \S 3.3.2.

Figure~5 shows the nearest neighbor
distributions for the LSBa and ORS spiral galaxies at declination  
$10<\delta <25^\circ$, which includes most of the 
\cite{LSBSample92} (1992) low and very low surface brightness
galaxies,
at galactic latitude $|b|>30^\circ$, and in the indicated ranges
of redshift corrected to the Local Group. To be
counted the nearest spiral neighbor must be more than 2~arc min
away, to reduce the chance the same object is in both catalogs, and  
at $cz>200$~km~s$^{-1}$, $0<\delta <35^\circ$, and
$|b|>30^\circ$. The cuts for LSBb (Fig.~6) are the
same except for the declination: the galaxies whose neighbors are 
counted are at $-1.5^\circ < \delta < 3.5^\circ$, and the spiral
neighbors are at $-11.5^\circ < \delta < 13.5^\circ$. The LSBb 
galaxies are the large angular size objects with HI redshifts in 
\cite{Impey96} (1996). (The optical redshifts  
are not used because their uncertainties would significantly add
to the nearest neighbor distances in redshift space.)

At the mean number density $n_o$ of spirals in the nearest
redshift bin, $200<cz<3000$ km~s$^{-1}$, the characteristic
physical distance to the nearest neighbor in the clustering
hierarchy is $H_oD_h\sim 35$ km~s$^{-1}$. As for the Dw/Irr case, 
this is small compared to typical relative peculiar velocities.  
The quite similar distributions of distances in redshift space to
the nearest spiral neighbors of these LSBs, 
Dw/Irr's, and spiral galaxies, imply both LSB and Dw/Irr
galaxies typically are closer than $1h^{-1}$~Mpc from a spiral
and all three galaxy types have quite similar relative peculiar
velocities. 

In the deepest redshift bin (the bottom line in Table~1) the 
mean nearest neighbor distance $\lb D_{ro}\rb$ for randomly
placed LSBa's is not significantly larger than the mean 
$\lb D_{to}\rb$ at the catalog positions. This can
be understood as follows. In a homogeneous random Poisson 
process with the mean number density $n_o=0.0007h^3$ Mpc$^{-3}$
of this sample of spirals the mean distance to the nearest
neighbor is    
\beq
D_{\rm Poisson} = {\Gamma(1/3)\over (36\pi n)^{1/3}}
\sim 6h^{-1}\hbox{ Mpc}. 
\label{eq:Poisson}
\eeq
This is comparable to $\lb D_{oo}\rb $ and to the clustering 
length $r_o$. This means the sparse sampling has suppressed the  
sensitivity of the nearest neighbor statistic to the galaxy
distribution. Figure~6 for LSBb
accordingly shows only the two 
lower redshift bins. 

At the number density of spirals in the bin $3000 < cz < 6000$
km~s$^{-1}$ the physical distance $D_h\sim 150$ km~s$^{-1}$ in
the clustering hierarchy is not much less than the relative
velocity dispersion. The physical separations thus might be
expected to make a significant contribution to the nearest
neighbor distances in redshift space. Consistent with this, 
$\lb D_{oo}\rb$ is not much larger than $D_h$.

The dotted lines in Figures~5 and~6 show the effect of randomly 
moving one in three of the LSB galaxies, leaving the rest at
their catalog positions. The effect is marginal in some cases,
but reproducible enough to show that fewer than one in three of
the LSBs could be in a statistically homogeneously distributed
population. Since most of this hypothetical homogeneous
population would be in voids I conclude that 

(4) the number density of observed LSB galaxies satisfies
\beq
n_{\rm LSB}\la 0.001h^3\hbox{ Mpc}^{-3}.
\label{eq:nlsb}
\eeq

\cite{LSB93} (1993) find that the nearest spiral neighbor of
an LSB is on average 1.7 times further than that of a spiral 
galaxy. The difference from the results presented here is a
cautionary example of the sensitivity to samples and methods of
analysis.

Bearing in mind this example, but considering also the
consistency of the distributions in Figures~5 and~6 from two
independent low surface brightness samples and two redshift bins,
I conclude that  

(5) the mean nearest neighbor distances in redshift space for
spirals and LSBs are not likely to differ by more than about 30
percent. 

Since this analysis uses low surface brightness galaxies detected
in HI the result is in line with the HI surveys in emission and
absorption that show that gas clouds very distinctly prefer to be
near galaxies (\S 2.3). But the quantitative constraints may be
useful. 

\subsection{Summary of the Nearest Neighbor Measure}

The candidate ordinary optically selected galaxies in a
simulation of galaxy formation ought to define realistic
voids. If the mean nearest neighbor distance among the candidate 
galaxies scales according to equation~(\ref{eq:Dh}), 
\beq
hD_h\sim 0.6\left( 0.01h^3\over n_o\hbox{ Mpc}^3\right) ^{0.8}
	\hbox{ Mpc},
\label{eq:Dc_value}
\eeq
when the cut in luminosity function or circular velocity produces
mean number density $n_o$, the distribution may be expected
to approximate the observed scale-invariant clustering hierarchy.
In this case the voids likely are satisfactorily large.

Candidate void galaxies in a simulation of galaxy formation are
the mass concentrations that are not good candidates for ordinary
galaxies --- perhaps their circular velocities are too small ---
but seem to be capable of developing into observable gas clouds
or galaxies of stars. They might be candidates for Magellanic-type
irregulars, with circular velocities greater than about $v_c\sim
20$~km~s$^{-1}$, large enough to resist substantial loss of
photoionized plasma (\cite{Rees86} 1986; \cite{Babul92} 1992;
\cite{Efstathiou92} 1992). Or they might be candidate low surface
brightness or compact galaxies. 

In the examples in \S 3.3 the mean physical distance from a
Dw/Irr or LSB galaxy to the nearest spiral satisfies   
\beq
\lb D_{to}\rb _{\rm physical}\la 1h^{-1}\hbox{ Mpc}, 
\label{eq:nntest}
\eeq
when the mean number density of spirals is
\beq
n_o\ga 0.01h^3\hbox{ Mpc}^{-3}.
\eeq
The samples avoid rich clusters, that would increase the mean 
distances in redshift space and reduce them in real space. 
In redshift space equation~(\ref{eq:nntest}) is replaced with an 
equality to about 10 percent. 

At smaller $n_o$ the examples in \S 3.3 indicate 
\beq
\lb D_{to}\rb _{\rm physical}\la 2\lb D_{oo}\rb _{\rm physical} 
	\sim 2D_h.
\label{eq:7}
\eeq
The factor of two takes account of applications of the
nearest neighbor test that find $\lb D_{to}\rb > \lb D_{oo}\rb$. 
In the examples presented here $\lb D_{to}\rb$ and 
$\lb D_{oo}\rb$ are equal to $\sim 20$ percent. 

These numbers can be considered a quantitative definition of 
the void phenomenon. One must add the cosmic number
density, $n_t$, because a model may satisfy
equation~(\ref{eq:nntest}) with a high density of void galaxies 
and an unacceptably high density of clustered objects.
Equation~(\ref{eq:nlsb}) gives a limit on the density
of candidate void LSB galaxies; the corresponding limit on void
dwarf plus irregular galaxies is
\beq
n_{\rm Dw/Irr}\la 0.01h^3\hbox{ Mpc}^{-3}.
\eeq

More observational checks of these numbers would be of interest.
The application to realizations of models for galaxy formation
might interesting too.   

\section{Theoretical Situation}

The conventional theoretical interpretation of voids was driven
by the elegance of the Einstein-de~Sitter model (with $\Omega _m=1$ 
in matter capable of clustering, and negligibly small space
curvature and cosmological constant), for this cosmological model
requires most of the mass to be in the voids. It also is motivated
by the CDM model for structure formation, which naturally
produces a biased distribution of galaxies relative to mass. The
considerations prior to the paradigm shift to $\Omega _m=0.25\pm 0.1$ 
are worth reviewing as a guide to the present situation.   

\subsection{Voids in an Einstein-de Sitter Universe}

The small relative velocity dispersion in the CfA sample
(\cite{DavisPeeb83} 1983) shows that if $\Omega _m=1$ then most
of the mass has to be in the voids.\footnote{The
relative velocity dispersion in the CfA sample is biased low by
the under-representation of rich clusters with large velocity
dispersions (\cite{Marzke95} 1995). But the mass within the Abell
radii of the Abell clusters is only about one percent of the
critical Einstein-de Sitter value, and the low relative
velocities of galaxies outside the clusters indicates the mass in
the less dense galaxy concentrations sampled by CfA is
significantly less than critical too.} \cite{DEFW85} (1985;
following \cite{Kaiser84} 1984 and \cite{Bardeen86} 1986) show this
could occur in a natural way in the CDM model: if ordinary galaxies
formed preferentially in high density regions they would be
strongly clustered, leaving most of the mass in the voids. The
void mass would be clumpy and thus might be expected to produce
bound objects, some observable though different from ordinary
galaxies. This seems intuitively reasonable (\cite{Peebles86}
1986, 1989). It is made quantitative in computations by
\cite{Einasto84} (1984), \cite{DS86} (1986),
\cite{Brainerd92} (1992) and \cite{Hoffman92}
(1992).\footnote{The cosmic string picture with hot dark matter
(\cite{Scherrer89} 1989), and the explosion variant
(\cite{Explosion86} 1986), might produce voids that 
never were substantially disturbed, though that does not
agree with the recent picture of a near space-filling  
Lyman-$\alpha$ forest at redshift $z\sim 3$.} 

Opinions on whether the distributions of galaxy types argue for 
or against the CDM model are mixed. Arguments in favor cite
morphological  segregation (\cite{DS86} 1986), the smaller
two-point correlation function for classes of unusual galaxies
(\cite{Salzer90} 1990; \cite{Mo94} 1994), and low surface
brightness galaxies (\cite{Hoffman92} 1992). Section 3 presents
reasons for treating these lines of evidence with some caution. 
Others conclude that the void phenomenon challenges
biased galaxy formation (\cite{Einasto84} 1984, 
\cite{Thuan87} 1987;
\cite{Eder89} 1989; \cite{Peebles89} 1989; \cite{Binggeli90}
1990; and \cite{Einasto94} 1994). 

\cite{Ostriker93} (1993) gives a balanced summary of the general
opinion in the community shortly before the paradigm shift to low
$\Omega _m$: ``Nominally  one expects, in the CDM model, that the 
voids will be populated to a degree larger than is observed ...
but in the absence of agreed-upon theories of galaxy formation,
it is difficult to quantify this apparent disagreement.'' 

Ostriker's assessment is accurate, but we do have guidance
on what might have happened from what is observed
(\cite{Peebles89} 1989). Consider the 
Magellanic-type irregulars on the outskirts of the Local Group
--- IC~1613, Sextans~A and~B, WLM, IC~5152, and
NGC~3109 --- at distances between 0.7 and 1.7~Mpc (\cite{Tully88} 
1988; Table~1 in \cite{Action00} 2000). They have small
peculiar velocities relative to the Local Group. They are not
near a large galaxy, so they are not likely to have been spawned
by tidal tails or other nonlinear process in the large galaxies.
Since they are at  ambient densities close to the cosmic mean
their first substantial star  
populations would have formed under conditions not greatly 
different from the voids at the same epoch. Why are such galaxies
so rare in the voids?

\subsection{Voids in a Low Density Universe}

If $\Omega _m=0.25\pm 0.1$ the observations are consistent with
the assumption that ordinary optically selected galaxies trace
the mass (\cite{Bahcall00} 2000 and references therein). If
galaxies are good mass tracers we can assume the voids contain
little mass. This is the natural interpretation of the void
phenomenon. But this consideration did not play a significant
role in the change of the most favored model from
Einstein-de~Sitter to low $\Omega _m$. I suspect one
reason is the prediction by CDM simulations that the voids
contain significant mass even when $\Omega _m$ is small. 

A visual impression of numerical simulations of the low density
$\Lambda$CDM model (with a cosmological constant to make flat
space sections) is that of classical biasing: the larger dark
mass halos cluster more strongly than the mass, and the less
massive halos spread into  
the voids defined by the larger halos. It is no criticism of
these studies, which report the behavior of the model, to
note that one is not reminded of the void phenomenon illustrated
in Figures~2 and~3. The situation is clearly presented in
Figure~5 of \cite{Kauffmann99} (1999).  

The remedy may be a proper understanding of how the formation of 
observable objects may be suppressed in voids 
(\cite{Ostriker93} 1993). \cite{Rees85} (1985)
explores the possible role of relativistic 
particles or ionizing radiation from the 
first generation of galaxies in suppressing subsequent galaxy 
formation, that otherwise would tend to occur in voids. 
\cite{Cen00} (2000) analyze realizations based on a numerical
prescription for how galaxies form or are suppressed. They
find the prescription produces observationally acceptable
realizations. An application of the nearest neighbor statistic in
\S 3.4 would be interesting. An explanation of the Im
phenomenon (\S 4.1) would be edifying.

The remedy may be an adjustment of the CDM model.
\cite{Bode00} (2000) show that warm dark matter can 
produce regions that are quite devoid of gravitational seeds for 
structure formation. The effect is striking, but the scale
unfavorable. If in their Figure~11
the large circles represented ordinary optically selected
galaxies, which would make about the observed number density, and
the small circles were dwarfs or irregulars, the model
would not seem to agree with the nearest neighbor distribution in  
Figure~4, or the maps in Figures~2 and~3. If the scale were
enlarged so all circles represented ordinary $L\sim L_\ast$
galaxies, the voids would be well represented, but the
Lyman-$\alpha$ forest at $z=3$ would be a problem. 

\section{Summary Remarks}

\subsection{The Observational and Theoretical Situations}

1. Some galaxy types prefer dense regions (\cite {Hubble36}
1936), others the edges of voids (\cite{Markarian83} 1983). This
well-established morphology-density correlation
(\cite{Dressler80} 1980) seems to arise in a natural way
from hierarchical gravitational structure formation, as in the
CDM model (\cite{White87} 1987).  

2. All known galaxy types and most gas clouds are scarce outside 
the concentrations of ordinary galaxies. This void phenomenon
is discontinuous from the morphology-density correlation 
(\S 2). It appears to be a more challenging test of ideas on how
galaxies formed (\S 4.2).  

3. Two-point correlation functions are sensitive probes
of the morphology-density correlation but, I argue in \S 3.1, are
not readily interpreted measures of the void phenomenon.  

4. The advantage of the nearest-neighbor statistic is a
reasonably simple interpretation as a constraint on void
galaxies. This is widely recognized in the observational
community (\S 3.2). The summary in \S 3.4 may be useful for
testing simulations of galaxy formation.  

5. The void phenomenon is observed among an impressively
broad range of objects. The weight of this evidence naturally
is from more readily observable gas-rich and star-forming
objects, however. It may be significant that the considerations
of \cite{DS86} (1986) argue for early-type void objects. Further
observational tests for such objects are particularly desirable. 

6. The challenge to the $\Lambda$CDM model might be resolved by a
demonstration that the formation of observable void objects
really can be adequately suppressed. This approach is challenged
in turn by the Magellanic-type irregulars on the outskirts of the
Local Group, that seem to have formed in ambient conditions not
very different from the voids (\S 4.1).

7. The challenge might be resolved by adjusting the model for
structure formation. A perhaps desperate idea is that the voids
have been emptied by the gravitational growth of holes in the
mass distribution (\cite{Peebles82} 1982). The void phenomenon
seems striking enough to motivate a search for viable initial 
conditions for this picture.  

\subsection{Interpretations of Voids as an Example of the
Scientific Method}

The introduction of simulations of biased galaxy formation 
(\cite{DEFW85} 1985) was not inspired by or even obviously
consistent with the evidence that giant and dwarf galaxies have
quite similar distributions (\cite{KOSS81} 1981; \cite{CfA82}
1982). This follows an honorable tradition in cosmology and, I
suspect, other developing sciences. A strikingly successful
example in cosmology is \cite{Einstein17}'s (1918) cosmological
principle. It did not agree with what was then known, but it led
us to an aspect of physical reality.

The advances in ideas on structure formation since 1985 have not 
been seriously influenced by the void phenomenon. Again, this is
not unusual. Another example is the galaxy n-point correlation
functions. The near power law form of the observed two-point
function is a widely discussed test. The three-point function is
little noted in discussions of simulations, despite its importance 
in characterizing the small-scale galaxy distribution 
(\cite{SP77} 1977). This complex and sometimes weak interplay of
theory and practice in cosmology is well represented by 
\cite{Kuhn62}'s (1962) paradigms, with socially selected theories
and constraints. The weak interplay can be healthy. It allows a
concentrated study of a particular subset of ideas, that may
establish or eliminate them. It reduces the chance of distraction
by misread evidence. It also allows distraction by unprofitable
ideas, of course, but that is 
remedied when theory and practice mature and on occasion produce
crises that drive paradigm shifts to better approximations to reality. 

The CDM model is maturing, most dramatically in its success
in relating the power spectrum of the
thermal background radiation temperature to observationally
acceptable cosmological parameters (eg. \cite{Hu00} 2000). This
shows the CDM model likely is a good approximation to how
structure started forming on the length scales probed by the
measurements. 

The apparent inconsistency between the theory and observations of
void is striking enough to be classified as a crisis for
the CDM model. It may be resolved within the model, through a
demonstration of an acceptable theory of galaxy formation. Or it
may drive an adjustment of the model. 

\acknowledgements

I have benefitted from discussions with Jerry
Ostriker, Stacy McGaugh, Michael Strauss, Laird Thompson, Trinh
Thuan, Brent Tully, and Neil Turok. This work was supported in
part by the National Science Foundation.


\begin{thebibliography}{99}
\bibitem[Babul \&\ Rees]{Babul92}
	Babul, A. \&\ Rees, M. J. 1992, MNRAS, 255, 346
\bibitem[Bahcall et al.]{Bahcall00} 
        Bahcall, N. A., Cen, R., Dav\'e, R., Ostriker, J. P.
        \&\ Yu, Q. 2000, astro-ph/0002310
\bibitem[Bardeen]{Bardeen86} 
	Bardeen, J. M. 1986, in Inner Space Outer Space, eds. E. W.
	Kolb, M. S. Turner, D. Lindley, K. Olive \&\ D. Sekel
	(Chicago: the University of Ghicago Press), p. 212
\bibitem[Bergeron \&\ Boisse]{Bergeron91} 
	Bergeron, J. \&\ Boisse, P. 1991, AA, 243, 344
\bibitem[Binggeli, Tarenghi \&\ Sandage]{Binggeli90}
 	Binggeli, B., Tarenghi, M. \&\ Sandage, A. 1990, AA, 228, 42
\bibitem[Bode, Ostriker \&\ Turok]{Bode00}
	Bode, P., Ostriker, J. P. \&\ Turok, N. 2000,
	astro-ph/0010389
\bibitem[Bothun et al.]{LSB93} 
	Bothun, G. D., Schombert, J. M., Impey, C. D.,
	Sprayberry, D. \&\ McGaugh, S. S. 1993, AJ, 106, 530 
\bibitem[Brainerd \&\ Villumsen]{Brainerd92}
	Brainerd, T. G. \&\ Villumsen, J. V. 1992, ApJ, 394, 409
\bibitem[Carignan \&\ Freeman]{Carignan88} 
	Carignan, C. C. \&\ Freeman, K. C 1988, ApJ, 332, L33
\bibitem[Cen \&\ Ostriker]{Cen00}
	Cen, R. \&\ Ostriker, J. P. 2000, ApJ, 583, 83
\bibitem[Chincarini \&\ Rood]{CR76} 
	Chincarini, G. \&\ Rood, H. J. 1976, ApJ, 206, 30
\bibitem[Chincarini, Rood \&\ Thompson]{bridges81} 
	Chincarini, G., Rood, H. J. \&\ Thompson, L. A. 1981,
	ApJ, 249, L47
\bibitem[Davis et al.]{DEFW85} 
	Davis, M., Efstathiou, G., Frenk, C. S. \&\ White, S. D. M. 
	1985, ApJ, 292, 371
\bibitem[Davis et al.]{CfA82}
	Davis, M., Huchra, J., Latham, D. W. \&\ Tonry, J. 1982, 
	ApJ, 253, 423
\bibitem[Davis \&\ Peebles]{DavisPeeb83}
	Davis, M. \&\ Peebles. P. J. E. 1983, ApJ, 267, 465
\bibitem[Dekel \&\ Silk]{DS86}
	Dekel, A. \&\ Silk, J. 1986, ApJ, 303, 39
\bibitem[de Lapparent, Geller \&\ Huchra]{deL86}
	de Lapparent, V., Geller, M. J. \&\ Huchra, J. P. 1986, 
	ApJ, 302, L1
\bibitem[Dressler]{Dressler80} 
	Dressler, A. 1980, ApJ, 236, 351
\bibitem[Eder et el.]{Eder89}
	Eder, J. A., Schombert, J. M., Dekel, A. \&\ Oemler, A.
	1989, ApJ, 340, 29
\bibitem[Efstathiou]{Efstathiou92}
	Efstathiou, G. 1992, MNRAS, 256, 43P
\bibitem[Einasto et al.]{Einasto84} 
	Einasto, J., Klypin, A. A., Saar, E. \&\ Shandarin, S. F.
	1984, MNRAS, 206, 529
\bibitem[Einasto et al.]{Einasto94} 
	Einasto, J., Saar, E., Einasto, M., Freudling, W.
	\&\ Gramann, M. 1994, ApJ, 429, 465
\bibitem[Einstein]{Einstein17}
	Einstein, A. 1917, S.-B. Preuss. Akad. Wiss., 142
\bibitem[El-Ad \&\ Piran]{Piran00} 
	El-Ad, H. \&\ Piran, T. 2000, MNRAS, 313, 553
\bibitem[Gregory \&\ Thompson]{GT78}
	Gregory, S. A. \&\ Thompson, L. A. 1978, ApJ, 222, 784
\bibitem[Grogin \&\ Geller]{GroginGeller00} 
	Grogin, N. A. \&\ Geller, M. J. 2000, AJ, 119, 32
\bibitem[Hermit et al.]{twopoint96} 
	Hermit, S., Santiago, B. X., Lahav, O., Strauss, M. A,
	Davis, M., Dressler, A. \&\ Huchra, J. P. 1996, 
	MNRAS, 283, 709
\bibitem[Hoffman, Lu \&\ Salpeter]{HIa92}
	Hoffman, G. L., Lu, N. Y. \&\ Salpeter, E. E. 1992, AJ,
	104, 2086
\bibitem[Hoffman, Silk \&\ Wyse]{Hoffman92}
	Hoffman, Y., Silk, J. \&\ Wyse, R. F. G. 1992, ApJ, 388, L13
\bibitem[Hu et al.]{Hu00}
	Hu, W., Fukugita, M., Zaldarriaga, M., \&\ Tegmark, M.
	2000, astro-ph/0006436 
\bibitem[Hubble]{Hubble36} 
	Hubble, E. 1936, Realm of the Nebulae (New Haven: Yale
	University Press)
\bibitem[Impey et al.]{Impey96} 
	Impey, C. D., Sprayberry, D., Irwin. M. J. 
	\&\ Bothun, G. D. 1996, ApJS, 105, 209
\bibitem[J\^oeveer, Einasto, \&\ Tago]{Ei78} 
	J\^oeveer, M., Einasto, J. \&\ Tago, E. 1978, MNRAS, 185, 357
\bibitem[Kaiser]{Kaiser84}
	Kaiser, N. 1984, ApJ, 284, L9
\bibitem[Kauffmann et al.]{Kauffmann99}
	Kauffmann, G., Colberg, J. M., Diaferio, A. \&\ White, S.
	D. M. 1999, MNRAS, 303, 188
\bibitem[Kirshner et al.]{KOSS81} 
	Kirshner, R. P., Oemler, A., Schechter, P. L. \&\
	Shectman, S. A. 1981, ApJ 248, L57 
\bibitem[Kuhn, Hopp \&\ Els\"asser]{Kuhn97} 
	Kuhn, B., Hopp, U. \&\ Els\"asser, H. 1997, AA, 318, 405
\bibitem[Kuhn]{Kuhn62}
	Kuhn, T. S. 1962, The Structure of Scientific
	Revolutions (Chicago: University of Chicago Press)
\bibitem[Lanzetta et al.]{Lanzetta95}
	Lanzetta, K. M., Bowen, D. V., Tytler, D. \&\ Webb,
	J. K. 1995, ApJ, 442, 538
\bibitem[Lee et al.]{ELG00} 
	Lee, J. C., Salzer, J. J., Rosenberg, J. L. \&\ Law, D.
	A. 2000, ApJ, 536, 606
\bibitem[Linder et al.]{BCG96} 
	Lindner, U. et al. 1996, AA, 314, 1
\bibitem[Marzke et al.]{Marzke95} 
	Marzke, R. O., Geller, M. J. da Costa, L. N.
	\&\ Huchra, J. P. 1995, AJ, 110, 477
\bibitem[Mayall]{Mayall60}
	Mayall, N. U. 1960, Annales d'Astrophysique, 23, 344
\bibitem[Mo, McGaugh \&\ Bothun]{Mo94} 
	Mo, H. J., McGaugh, S. S. \&\ Bothun, G. D. 1994, 
	MNRAS, 267, 129
\bibitem[Ostriker]{Ostriker93}
	Ostriker, J. P. 1993, Ann. Rev. Astron. Ap., 
	31, 689
\bibitem[Ostriker, Thompson \&\ Witten]{Explosion86}
	Ostriker, J. P. Thompson, C. \&\ Witten, E. 1986, Phys.
	Letters B, 180, 231
\bibitem[Peebles]{Peebles82} 
	Peebles, P. J. E. 1982, ApJ, 257, 438
\bibitem[Peebles]{Peebles86} 
	Peebles, P. J. E. 1986, Nature, 321, 27
\bibitem[Peebles]{Peebles89} 
	Peebles, P. J. E. 1989, J. Roy. Astron. Soc. Canada 1989,
	83, 363
\bibitem[Peebles]{Pee93} 
	Peebles, P. J. E. 1993, Principles of Physical Cosmology
	(Princeton: Princeton University Press)  
\bibitem[Peebles et al.]{Action00}
	Peebles, P. J. E., Phelps, S. D., Shaya, E. J. \&\
	Tully, R. B. 2000, astro-ph/0010480
\bibitem[Popescu, Hopp \&\ Els\"asser]{Popescu97}
	Popescu, C. C. Hopp, U. \&\ Els\"asser, H. 1997,
	AA, 325, 881
\bibitem[Postman \&\ Geller]{Postman84}
	Postman, M. \&\ Geller, M. J. 1984, ApJ, 281, 95
\bibitem[Pustil'nik et al.]{Pustilnik95}
	Pustil'nik, S. A., Ugryumov, A. V., Lipovetsky, V. A. 
	Thuan, T. X. \&\ Guseva, N. G. 1995, ApJ, 443, 499
\bibitem[Pustil'nik et al.]{Pustilnik00} 
	Pustil'nik, S. A., Brinks, E., Thuan, T. X., Lipovetsky,
	V. A. \&\ Izotov, Y. I. 2000, astro-ph/0011291
\bibitem[Rees]{Rees85} 
	Rees, M. J. 1985, MNRAS, 213, 75P
\bibitem[Rees]{Rees86} 
	Rees, M. J. 1986, MNRAS, 218, 25P
\bibitem[Rood]{Rood81} 
	Rood, H. J. 1981, Rept. Prog. Phys., 44, 1077
\bibitem[Salzer] {Salzer89} 
	Salzer, J. J. 1989, ApJ, 347, 152
\bibitem[Salzer, Hanson, \&\ Gavazzi]{Salzer90} 
	Salzer, J. J., Hanson, M. M. \&\ Gavazzi, G. 1990, ApJ, 353, 39
\bibitem[Santiago et al.]{Santiago95}
	Santiago, B. X., Strauss, M. A., Lahav, O., Davis, M., 
	Dressler, A.  \&\ Huchra, J. P. 1995, ApJ, 446, 457
\bibitem[Scherrer et al.]{Scherrer89} 
	Scherrer, R. J., Melott, A. L. \&\ Bertschinger, E.
	1989, PRL, 62, 379 
\bibitem[Schombert et al.]{LSBSample92} 
	Schombert, J. M., Bothun, G. D., Schneider, S. E. \&\
	McGaugh, S. S. 1992, AJ, 103, 1107
\bibitem[Schombert, Pildis \&\ Eder]{Schombert97}
	Schombert, J. M., Pildis, R. A. \&\ Eder, J. A. 1997,
	ApJS, 111, 233
\bibitem[Shull, Stocke \&\ Penton]{Shull96}
	Shull, J. M., Stocke, J. T. \&\ Penton, S. 1996, AJ, 111, 72
\bibitem[Soneira]{Soneira78} 
	Soneira, R. M. 1978, PhD dissertation, Princeton
	University
\bibitem[Soneira \&\ Peebles]{SP77} 
	Soneira, R. M. \&\ Peebles, P. J. E. 1977, ApJ, 211, 1
\bibitem[Soneira \&\ Peebles]{SP78} 
	Soneira, R. M. \&\ Peebles, P. J. E. 1978, AJ, 83, 845
\bibitem[Steidel, Dickinson \&\ Persson]{Steidel94}
	Steidel, C. C., Dickinson, M., \&\ Persson, S. E.
	1994, ApJ, 427, L75
\bibitem[Szomoru et al.]{HIc96}
	Szomoru, A., van Gorkom, J. H., Gregg, M. D. \&\ Strauss,
	M. A. 1996, AJ, 111, 2150
\bibitem[Thompson]{Markarian83}
	Thompson, L. A. 1983, ApJ, 266, 446
\bibitem[Thuan, Gott \&\ Schneider]{Thuan87} 
	Thuan, T. X., Gott, J. R. \&\ Schneider, S. E. 1987,
	ApJ, 315, L93
\bibitem[Thuan \&\ Izotov]{Thuan97}
	Thuan, T. X. \&\ Izotov, Y. I. 1997, ApJ, 489, 623
\bibitem[Tifft \& Gregory]{TG76} 
	Tifft, W. G. \&\ Gregory, S. A. 1976, ApJ, 205, 696
\bibitem[Tully]{Tully88}
	Tully, R. B. 1988, Nearby Galaxies Catalog (Cambridge:
	Cambridge University Press)
\bibitem[Vanzi et al.]{Vanzi00}
	Vanzi, L., Hunt, L. K., Thuan, T. X. \&\ Izotov, Y. I.
	2000, astro-ph/0009218
\bibitem[Vettolani et al.]{Vett85}
	Vettolani, G., de Souza, R. E., Marano, B. \&\
	Chincarini, G. 1985, AA, 144, 506
\bibitem[Vogeley et al.]{Vogeley94} 
	Vogeley, M. S., Geller, M. J., Park, C. \&\ Huchra, J. P.
	1994, AJ, 108, 745 
\bibitem[Weinberg et al.]{HIb91}
	Weinberg, D. H., Szomoru, A., Guhathakurta, P. \&\ van
	Gorkom, J. H. 1991, ApJ, 372, L13
\bibitem[White]{White79} 
	White, S. D. M. 1979, MNRAS, 186, 145
\bibitem[White et al.]{White87} 
	White, S. D. M., Davis, M., Efstathiou, G. \&\ 
	Frenk, C. S. 1987, Nature, 330, 451
\bibitem[Willick et al.]{Willik97}
	Willick, J. A., Courteau, S., Faber, S. M., Burstein, D.,
	Dekel, A. \&\ Strauss, M. A. 1997, ApJ Suppl, 109, 333
\bibitem[Zwaan et al.]{Zwaan97}
	Zwaan, M. A., Briggs, F. H., Sprayberry, D. \&\
	Sorar, E. 1997, ApJ, 490, 173
\end{thebibliography}
\end{document}